
%
%
%
\documentstyle{amsppt}
\nologo

\hsize 32pc
\vsize 50pc
\emergencystretch=100pt

\magnification =1200
\pageheight{ 7.5 in}
\baselineskip=25pt plus 2pt

%
%

\def\ms{{\medskip}}

\def\sla{{\Cal A}}
\def\slk{{\Cal K}}
\def\sle{{\Cal E}}
\def\sln{{\cal N}}
\def\sln{{\Cal N}}
\def\slr{{\Cal R}}
\def\sll{{\Cal L}}
\def\slh{{\Cal H}}

\def\slf{{\Cal F}}

\def\Tau{{\Cal T}}
\def\slp{{\Cal P}}

\def\slz{{\Cal Z}}

\def\lb{{\{}}
\def\rb{{\}}}

%
%

\topmatter

\title
{The localized induction equation,} \\
 {the Heisenberg chain, }\\
{and the} \\
 {non-linear Schr\"odinger equation}
\endtitle
\author
 Joel Langer  and  Ron Perline
\endauthor
\affil
Dept. of Mathematics, Case Western Reserve University \\
Dept. of Mathematics and Computer Science, Drexel University
\endaffil
\abstract
The three equations named in the title are examples
of infinite-dimensional completely integrable Hamiltonian systems,
and are related to each other via  simple geometric
constructions.  In this paper, these interrelationships are further explained
in terms of the recursion operator for the Localized Induction Equation,
and the recursion operator is seen to play a variety of roles in key
geometric variational formulas.
\endabstract
\endtopmatter

\rightheadtext{(LIE), (HC), and (NLS)}

{\bf Remarks:  The authors would appreciate it if anyone who downloads
this file from the nonlinear science preprint archives would send an
acknowledgement to: rperline\@mcs.drexel.edu.  We are interested in tracking
the archive usage.  Thanks!}
\ms
\ms
{\bf 0. INTRODUCTION} \newline
Among the natural variational integrals in geometry are the
{\it inevitable integrals} on space curves $\gamma(s)$; these
include {\it length} $\sll(\gamma) = \int ds $, {\it total torsion} $
\Tau(\gamma) = \int \tau ds$, {\it total squared curvature} $ \slk(\gamma) = {1
\over 2}
\int \kappa ^2 ds$, the integral
 $ \slf(\gamma) = {1 \over 2} \int \kappa^2 \tau ds \ = \
{1 \over 2} \int \gamma_s \times \gamma_{ss} \cdot \gamma_{sss}\  ds$,
and an infinite sequence of integrals following these four.  A closely
related sequence, for curves $T( \sigma)$ on the sphere, begins with
{\it total geodesic curvature} $ \sla(T) = \int \kappa _T ds_T$, and {\it
energy } $ \sle(T) = \int {1 \over 2} |  T_{\sigma} |^2 d \sigma$.
(By the Gauss-Bonnet theorem we can view $\sla(T)$ as the {\it spherical
area}
bounded by $T$, when $T$ is  closed).  Though the whimsical term
{\it inevitable} will be explained here by some simple geometric
considerations,
these sequences first arose via some physical models (which the geometric
point of view has helped illuminate).

In fact, the length integral $\sll$ may be regarded as the {\it
Hamiltonian}
for a model of thin vortex tubes in three dimensional hydrodynamics, where
the evolution of the tube's centerline $\gamma (s,t)$
is governed by the {\it Localized Induction Equation} (LIE):
$\gamma_t = \gamma_s \times \gamma_{ss} = \kappa B$ (see [M-W]).
Then the remaining
integrals in the first list are {\it constants of motion} for the
evolving curves.  Similarly, the energy $\sle$ is the Hamiltonian for the
(continuous) {\it Heisenberg Chain} (HC) (see [F-T]), $T_t = T \times T_{\sigma
\sigma}$,
while $\sla$ and the integrals following $\sle$ are constants of motion for
the
curves $T(\sigma, t)$ evolving on the sphere.

The simple geometric relationship between these two {\it soliton equations}
is this : if $\gamma$ satisfies (LIE), then the {\it tangent indicatrix of
$\gamma, \ T = \gamma_s$}, satisfies (HC).  By the way, it pays not to
check
this fact too hastily.  After differentiating the first equation with
respect
to $s$, one wants to replace $\gamma_{ts}$ by $\gamma_{st} = T_t$ to get
the
left hand side of (HC).  To justify this one has to observe that
$W = \kappa B$ happens to belong to the special class of {\it locally
arclength preserving} (LAP) variation vectorfields.  $W$ is LAP if the
speed
along $\gamma$ is preserved under the evolution $\gamma_t =W$ -- in
particular,
an arclength parameter $s$ remains an arclength parameter -- which is
equivalent to the condition that $W_s$ is orthogonal to $T$ (obviously
satisfied by $W = \kappa B = T \times T_s$).

The two  lists of constants  are also  related by the tangent indicatrix
construction;
if $T = \gamma_s$, then $\Tau(\gamma )
 = \sla (T), \slk (\gamma) = \sle (T), $ etc.
(Note that the second list is `missing' one integral, namely   $\sll$.) The
two lists are essentially `equivalent', but the first
is of perhaps greater interest in the study of
{\it geometric variational problems}.  For
example, one of the oldest problems in the calculus of variations is
that of the {\it Bernoulli elastic rod}, to `minimize' $\slk$ with
a constraint on $\sll$.  More generally, the extrema for sums
$\alpha_1 \sll + \alpha_2 \Tau +\alpha_3 \slk$, $\alpha_i$ constants, are
precisely the (centerlines of) {\it Kirchhoff elastic rods} in equilibrium.
(Actually, this simple characterization seems to be relatively unknown; the
{\it twist energy}, which the Kirchhoff model adds to the Bernoulli model
is
given by a quadratic integral on $framed$ curves [L-S]).  To our knowledge, no
one
has classified  the extrema for sums
$\alpha_1 \sll + \alpha_2 \Tau +\alpha_3 \slk + \alpha_4 \slf$.

The first of these variational problems was introduced into the study
of (LIE) in 1971 by Hasimoto, who observed that if $\gamma(s,0)$ represents
a (Bernoulli) elastic rod in equilibrium, then $\gamma$ moves through
space under (LIE)    without changing shape [Has 1].  More generally (not part
of Hasimoto's original observation) such {\it one-soliton solutions} to
(LIE) are precisely the equilibria for the Kirchhoff model.

With this beginning, Hasimoto went on to relate the filament equation
itself to yet another physical model -- already known to exhibit
soliton behavior -- the {\it Non-linear Schr\"odinger Equation} (NLS)
: $ -i \psi_t = (\psi _{ss} + {1 \over 2}|\psi|^2 \psi) $.  Hasimoto showed
that if a
space curve $\gamma$ evolves according to (LIE), then the {\it curvature}
and {\it torsion} of $\gamma$, $\kappa(s,t) \ \hbox{and} \ \tau(s,t)$, can
be combined into a {\it complex curvature function} $\psi(s,t) =
\kappa e ^{i \int \tau ds}$, which satisfies (NLS) [Has 2].  The proof of this
fact
amounts to computing variational  formulas for $\kappa$ and $\tau$ or
equivalently, for the complex curvature $\psi$.  It turns out that these
formulas have a special structure, which is related to the infinite list
of constants of motion for (LIE), and which also clarifies the precise
nature of the equivalence of the three models.
The study of these formulas and their relatives,
their geometric interpretation, and their geometric
applications form the central theme of this paper.

To describe this special structure, let $\slh (\gamma) = \psi$ denote the
{\it Hasimoto transformation} taking space curves to complex curvature
functions.  Then the desired variational formula for $\psi$ is contained in
the
following  {\it general formula for the differential of $\slh$
at $\gamma$
in the direction of $W$, a LAP vectorfield}:
$$ d  \slh (W)   \equiv -\slz \slr^2 (W) .  \leqno{(1)}$$
The most important feature of this formula (whose proof is included below)
is the appearance of the (squared) {\it recursion operator} $\slr$, a
linear
integro-differential operator on vectorfields.  The operator $\slz$ can be
thought of as a simple isomorphism from LAP vectorfields to complex
functions of $s$; formulas for $\slr$ and $\slz$ will be given below.
We write $ \equiv $ to denote equivalence modulo addition of terms of the
form
$i c \psi = i c \slh (\gamma), c$ a real constant. Note we  have defined
$\slh$
itself only up to multiplicative factors of the form $e^{i \theta}$,
$\theta$ a `constant of integration' -- this ambiguity is not reflected
in $\kappa$ or $\tau$.  For  a varying curve, differentiation of
$\theta = \theta(t)$ produces an extra term $ic \psi$ in the
variation of $\psi$.  Setting $W= \kappa B$ in (1) easily yields
$\psi _t = d \slh (W) = i(\psi_{ss} +{1 \over 2}(|\psi|^2 +c(t) )) \psi )$,
which
is the precise statement of Hasimoto's result.
\ms

{\bf 1. THE RECURSION OPERATOR AND NATURAL FRAMES} \newline
To discuss formulas for $\slr$ and $\slz$ we use {\it natural frames}
along $\gamma$, in place of the standard {\it Frenet frame}
$\slf = \lb T,N,B \rb $.  Recall the {\it Frenet equations} may be written
in the form $T_s = \Omega \times T, N_s = \Omega \times N, B_s = \Omega
\times
B$, where $\Omega$ is the {\it Darboux vector} ({\it s-angular velocity}
of $\slf$) given by $\omega = \tau T + \kappa B$.  Similarly, the {\it
natural frames} are those orthonormal frames $\sln = \lb T, U , V \rb$
along
$\gamma$ having {\it s-angular velocity}
$\omega ^s = -vU + uV$, for some `curvature functions'
$u = u(s)$ and $ v = v(s)$; so $\sln$ satisfies the `Frenet Equations'
$T_s = \omega^s \times T = uU +vV, U_s = -uT, V_s = -vT$ . Note that
$\omega^s = T \times T_s = \kappa B$, the vectorfield defining (LIE)!   In
other words, among all ({\it adapted} orthonormal) frames
$\sln = \lb T, U, V \rb$, the natural frames are characterized by:
$\sln$ has zero tangential component of angular velocity . One might prefer
to call $\sln$ natural if it has  constant $T$-component of
angular velocity -- think of a bead sliding and turning without friction
along a wire $\gamma$ -- but we will not use this more
general class here.

An alternate way to describe  natural frames is to say that the
vectors  $ \lb U, V \rb$ normal to $\gamma$ are obtained by
{\it parallel translation in the sphere} along the tangent indicatrix
$T$.  Note that while $\slf$ is uniquely determined, $\sln$ is determined
only after an `initial frame' has been chosen at some point along $T$.
An advantage of $\sln$ (which plays no role presently) is that $\sln$ can
be defined even where $\gamma$ has vanishing curvature.

We  give formulas for $\slz (W)$ and $\slr (W)$, where
$W = aT + bU + cV$, and we also provide a simpler expression for the
complex curvature function $\psi$:

$$
\eqalign{
&\psi = \slh (\gamma) = u + iv, \cr
&\slz (W) = b + ic, \cr
&\slr (W) = - \slp (T \times W_s), \cr
&\slp (W) = (\int {bu +cv} \, ds ) T + bU + cV} \leqno{(2)}
$$

Here we have also introduced a {\it parameterization operator $\slp$},
which
leaves the normal part of an arbitrary vectorfield $W$ alone but turns
$W$ into a LAP vectorfield (since ${ d \over {ds}}
\slp (W)$ has no $T$  component).
It follows that the recursion operator $\slr$ preserves the class of LAP
vectorfields.  Note that we are again being casual about constants of
integration; $\slp (W)$ has only been defined up to addition of terms of
the
form ${constant} \cdot T$, and by the same token, $\slz$ is not quite
an isomorphism of LAP vectorfields.

Until now, we haven't said much about the meaning of the recursion operator
$\slr$ --  we have only  indicated
that it allows for a very compact expression for the
variation of the complex curvature.  To get some experience with
$\slr$, let's apply it to the humblest of all LAP vectorfields:
$\slr (T) = -\slp (T \times T_s) = -\slp (\kappa B) = -\kappa  B$, an
old friend! In this computation we simply ignored $\slp$ since
$\kappa B$ is already LAP.

Since this worked out so nicely, let's apply $\slr$ again (keeping in mind
that tangential terms inside of $\slp (\,)$ can be discarded):
$\slr^2 (T) = \slr(-\kappa B) = \slp (T \times (T \times T_s)_s) =
\slp (-T_{ss}) = -\slp (u_s U + v_s V) =
-(\int {u_s u + v_s v} ds)T -u_s U -v_s V = - {1 \over 2} (u^2 +v^2)T
-u_s U -v_s V$.  Note our good fortune in being able to compute the
antiderivative explicitly!  As it turns out, this will continue to
happen as we successively apply $\slr$, and the LAP vectorfields
$X_n = \slr^n (-T), n=1,2,3, \dots, $ are none other than the Hamiltonian
vectorfields of the constants of motion for the Localized Induction Equation
--
hence the  names ``recursion operator''  and  ``inevitable integrals''.
\ms
{\bf 2. THE VARIATIONAL  FORMULAS} \newline
We have explained the name, but only part of the significance
of $\slr$.  Recall that the formula (1) relates variations in the
`first' model (LIE) to variations in the `third' model (NLS).  Since
the {\it square} of $\slr$ occurs in this formula, it seems reasonable
that a  first power of $\slr$ ought to relate the `second' model
(HC) -- this is clearly the  intermediate model -- to (LIE) and (NLS).
To see how this is true, it is useful to modify the (HC) model slightly,
replacing the curve $T$ by a natural frame $\sln = \lb T , U, V \rb$.  Note
that in doing so, we have only enlarged our `state space' by one
dimension, corresponding to our usual ambiguity (say, the choice of
`initial frame' $\lb U_0 , V_0 \rb$, or choice of `phase factor'
$e^{i \theta} $ for $ \psi$).

Now suppose $\sln (s,t)$ evolves in time according to
$T_t = \omega^t \times T, U_t = \omega ^t
 \times U, V_t = \omega^t \times V$, for
some {\it t-angular velocity}
$\omega^t = \omega^t (s,t)$.  For $\sln$ to represent an
 evolving natural frame, it must satisfy
$\ 0 = U_s \cdot V$, for all $s,t$.  It follows that for $\omega^t$ to be
{\it naturality-preserving} (NP) it must satisfy:
$ 0 = (U_s \cdot V)_t = U_{ts} \cdot V + U_s \cdot V_t =
(\omega^t \times U)_s \cdot V +  \mathbreak U_s \cdot (\omega ^t \times V)
=
(\omega ^t)_s \times U \cdot V = (\omega ^t)_s \cdot T$, a familar
condition! In other words, allowing the same vectorfield
$( aT +bU +cV)$ to play two different roles:
{\it $W = aT +bU +cV$ is LAP if and only if $\, \omega ^t = aT +bU + cV$ is
NP.}
In particular, this means that if $\omega ^t$ is NP, so is
$\slr ^n (\omega ^t), n=1,2,3, \dots .$

Without further ado, we state the promised relationship between the
variations of curves and variations of the corresponding natural frames:
{\it If $W$ is LAP, and $\gamma(s,t)$ evolves according to
$\gamma _t = W$, then the $t$-angular velocity of a natural frame
$\sln = \lb T,U,V \rb$ along $\gamma$  is given by }
$$\omega ^t = -\slr W \leqno{(3)}$$
Proof:  Since $W = aT +bU +cV$ is assumed to be LAP, we
have $\gamma_{st} = \gamma_{ts}$, so $\omega^t \times T = T_t =
\gamma_{ts} = (aT  + bU + cV)_s = (au + b_s)U + (a v + c_s) V$.  It
follows that $\omega ^t = \eta T -(av +c_s) U + (au +b_s)V$, for some
function $\eta = \eta (s,t)$.  To determine $\eta$, we compare expressions
for $U_{st}$ and $U_{ts}$:
$U_{st} = (-uT)_t = -u_t T -u(au +b_s)U -u(av +c_s)V$, while
$U_{ts} =(\omega^t \times U)_s = ((\eta T +(au +b_s)V) \times U)_s =
(-(au + b_s)T + \eta V)_s =
-(b_{ss} + a_su + au_s + \eta v)T -u(au +b_s)U +(\eta_s -auv -vb_s)V$, so
equating $V$-coefficients gives $\eta_s -b_s v = -u c_s$.  Therefore
$\omega^t = (\int {vb_s -uc_s } ds)T -(av +c_s)U +(au +b_s)V$.  Finally,
$-\slr (W) = \slp(T \times (aT +bU +cV)_s) = \slp( T\times ((au +b_s)U
+ (av +c_s)V) = \slp (-(av +c_s)U +(au +b_s)V) =
(\int { -(av +c_s)u + (au+b_s)v} ds) T -(av+c_s)U +(au +b_s)V = \omega^t$.

It remains to see how $\slr$ takes us from the second model to the
third model.  Given a complex function $\psi = \psi(s,t)$, we can always
write the $t$-derivative $\psi_t$ in the form $\psi_t = - \slz( \phi ^t)$,
for
some vector field $\phi^t = aT +bU +cV$.  Of course, $\phi^t$ is not unique
since $\slz$ ignores tangential terms.  But if $\phi^t $ is also required
to be LAP, then it {\it will} be unique up to the usual terms, $constant
\cdot T$.  Now the question is, if a natural frame $\sln = \lb T,U,V \rb$
evolves with $t$-angular velocity $\omega^t$, and if $\psi = u+iv =
\slz (uU +vV) = \slz (T_s)$ evolves accordingly, what is the vectorfield
$\phi^t$?  The required computation is simplified by the abuse of notation
$Z = U + iV, \slz(W) = \slz \cdot W$ (complex inner product):
$\psi_t = (\slz  \cdot T_s)_t = \slz_t \cdot T_s + \slz \cdot T_{ts} =
\omega^t \times \slz \cdot T_s + \slz \cdot (\omega^t \times T)_s
= \slz \cdot (-T \times (\omega^t)_s) = \slz (\slr (\omega^t))$.  In other
words, we have  the following
{\it formula  relating the variations in natural frames and complex wave
functions}:
$$ \phi^t = -\slr (\omega ^t). \leqno{(4)} $$
Finally, note that in case $\sln (s,t)$ comes from a curve $\gamma (s,t)$,
and $\gamma_t =W$ is LAP, we have $\psi_t = \slz ( \slr (\omega^t))
= \slz (\slr ( - \slr (W)) = -\slz \slr ^2 (W)$, which proves (1).

In order to summarize the various roles of the recursion operator $\slr$ we
have included a diagram below.  Note that $\slr$ connects {\it levels}
corresponding to the three models (LIE), (HC), and (NLS)  (for the
actual equations (LIE), (HC), and (NLS), set $n=1$ )  and $\slr$
also connects $t$ and $s$ {\it directions of motion}.  Some of the arrows
have not been explained, but are easily interpreted and checked.

$$
\CD
{\gamma_s}      	@>{- \slr ^n}>>   {\gamma_t} \\
@V{-\slr}VV          @VV{-\slr}V \\
{\omega^s}       @>{- \slr ^n}>> {\omega^t} \\
@V{-\slr}VV        @VV{-\slr}V \\
{\phi^s}       @>{- \slr ^n}>> {\phi^t} \\
\endCD
$$

\medskip
We have not discussed here the Hamiltonian nature of the equations (LIE),
(HC), or (NLS), or the actual definitions of the spaces on which these
Hamiltonian flows are defined.
For background on such points see [L-P 2], where (1) is first proved and
used to establish the precise nature of the Hasimoto transformation as
a {\it Poisson map}.  Arguments similar to those used in [L-P 2] can be
combined with formulas (3) and (4) to give similar characterizations
of the intermediate maps discussed here (thus interpreting $\slh$ as a
composite of Poisson maps).
However, our intent  here has been to emphasize the ubiquity of the
recursion operator
$\slr$ and its relation to geometric varational problems and
flows of geometric origin.


\Refs\nofrills{References}

\widestnumber\key{Ge-Di23}


\ref \key{\bf F-T} \by L. Faddeev and L. Takhtajan
\book Hamiltonian methods in the theory of solitons
\publ Springer-Verlag \publaddr Berlin \yr1980
\endref

\ref \key{\bf Has1} \manyby H. Hasimoto
\paper Motion of a vortex filament and its relation
 to elastica  \jour  J. Phys. Soc. Japan
\yr 1971 \vol 31 \page 293
\endref

\ref \key{\bf Has2} \bysame
 \paper A soliton on a vortex filament
\jour  J. Fluid Mech.
\yr 1972 \vol 51 \page 477
\endref

\ref \key{\bf La} \by G.L. Lamb
\book Elements of soliton theory \publ Wiley Interscience
\publaddr New York  \yr 1980
\endref

\ref \key{\bf L-P 1}
\manyby J. Langer and R. Perline
\paper  The Hasimoto transformation and integrable flows on curves
\jour Appl. Math. Lett.
\yr 1990 \vol 3(2) \page 61
\endref

\ref \key{\bf L-P 2} \bysame
\paper Poisson geometry of the Filament Equation
\jour J. Nonlinear Sci.
\year 1991 \vol 1 \page 71
\endref

\ref \key{\bf L-S} \by J.Langer and D. Singer
\paper Lagrangian aspects of the Kirchhoff elastic rod
\paperinfo preprint
\endref

\ref \key{\bf M-W} \by  J. Marsden and A. Weinstein
\paper Coadjoint orbits, vortices, and Clebsch variables for
incompressible fluids
\jour Physica 7D \yr1983 \page 305
\endref \ms


\endRefs
\enddocument